\documentstyle[12pt]{article}
\begin{document}
\begin{titlepage}
\title{Clifford Algebra of Nonrelativistic Phase Space\\ and the Concept of
Mass}
\author{
{P. \.Zenczykowski $^*$ }\\
{\em Division of Theoretical Physics},
{\em Institute of Nuclear Physics}\\
{\em Polish Academy of Sciences}\\
{\em Radzikowskiego 152,
31-342 Krak\'ow, Poland}\\
e-mail: piotr.zenczykowski@ifj.edu.pl
}
\maketitle
\begin{abstract}
Prompted by a recent demonstration that the  
structure of a single quark-lepton generation 
may be understood via a Dirac-like linearization of the form ${\bf p}^2+{\bf x}^2$, 
we analyze the corresponding Clifford algebra in some detail. 
After classifying
all elements of this algebra according to their $U(1) \otimes SU(3)$ and $SU(2)$ 
transformation properties, we identify the element 
which might be associated with the 
concept of lepton mass. This element is then transformed into a
corresponding element for a single coloured quark.
It is shown that - although none of the three thus obtained
 individual quark mass elements 
is rotationally
invariant -
the rotational invariance of the quark mass term
 is restored when the sum over quark colours is performed.
\end{abstract}

\end{titlepage}

\section{Introduction}

The present paper develops somewhat further the approach proposed
in  \cite{APPB2007}-\cite{Zen2006}.
The ideas and heuristic arguments which provide a conceptual justification for the
whole approach
were originally presented in \cite{Zen2006}. They stemmed from: 1) dissatisfaction with the way quark mass
is introduced and used in contemporary elementary particle physics, and 2) a wish to
introduce more symmetry between position and 
momentum.\footnote{ While these two arguments are as valid as ever, 
one may now
say with a hindsight  that the particular way
in which symmetry between position and  momentum 
was enforced in \cite{Zen2006} suffered from
 an erroneous use
of $U(1) \otimes SU(3)$ as a means of effecting the anticipated lepton-quark interchange. 
Such a use of $U(1)\otimes SU(3)$ was forced by the requirement imposed in \cite{Zen2006} 
that position-momentum Poisson
bracket be invariant.  
It appears now, as found in \cite{APPB2007b}, 
that the corresponding invariance of the
position-momentum commutation relations has to be
somewhat relaxed and admit arbitrary ($+$ or $-$) 
signs in front of the imaginary unit, independently for each 
of the three directions of our
macroscopic 3D world.
The lepton-quark interchange is then effected by transformations outside of
$U(1)\otimes SU(3)$.}
In ref.\cite{Zen2006} it was argued that instead of identifying the
arena of nonrelativistic physics with the observable three-dimensional space,  
one should adopt the description given by the nonrelativistic
Hamiltonian formalism,
in which momentum and position coordinates are treated as {independent}
variables. In this language, the relevant arena appears to be that of
phase space. As a result, 
the question of a possible symmetry between
the momentum and position coordinates (advocated long ago by Born \cite{Born1949})
may be then formulated in a more natural way. 
This redefinition of what constitutes an arena admits a generalization of the way 
in which ``canonical momenta"
and ``canonical positions" are identified with physical momenta and physical positions. 
Thus, it appears that instead of just one 
way suggested by the 3D formalism, one can perform 
such an identification in four ways (including the old 3D one). 
Ref. \cite{Zen2006} put also forward 
a conjecture concerning the generalization of the standard concept of mass. 
It consisted in associating the concept of mass not only with the physical momentum 
(as in standard relations between energy, mass and momentum),
but, more generally, with the very four types of
``canonical momenta".
It was then further suggested that the additional three ways of assigning the
concept of mass
are related to the existence
of quarks, and that it is the ensuing lack of rotational invariance 
which is connected with
both quark unobservability and the conceptual problems related to
the way quark mass is introduced
in contemporary physics.

In ref.\cite{APPB2007} the above idea was developed further
by admitting non-commuting
positions and momenta. Subsequent Dirac-like linearization of 
the basic phase-space invariant
$R^z \equiv {\bf p}^2+{\bf x}^2$ 
has led to the 
appearance of a corresponding matrix operator $R$ in Clifford 
algebra,\footnote{The superscript $\sigma$, used in \cite{APPB2007}-\cite{APPB2007b} 
to distinguish 
between the matrix and phase-space
representations, is suppressed throughout the present paper.}
and to the proposal of
its identification  (up to a factor) with the hypercharge operator.
Similarly, the operator $R^{tot}=R^z+R$, 
with the lowest ``vacuum" eigenvalue adopted for $R^z$,
was conjectured to be identical (up to a factor) to the charge operator $Q$.
The resulting set of the eigenvalues of $Q$ corresponds precisely to the set of 
quark and lepton charges of a single Standard Model (SM) generation.
At the same time, the study of the relevant Clifford algebra provided
a raison d'\^etre for the appearance of the symmetry group $U(1) \otimes SU(3)$ combined
with $SU(2)$, a conjectured precursor of the SM gauge group.
 Furthermore,
it was shown \cite{Zen2008,APPB2007b} that there is 
a one-to-one correspondence between 
the way in which charge eigenvalues emerge in the proposed scheme 
and in the Harari-Shupe rishon model
\cite{Harari}.
Finally, a phase-space-based interpretation of the connection between 
leptons and quarks was proposed, with quarks being related to leptons
via genuine rotations in phase space and 
weak isospin related to reflections in phase space \cite{Zen2008,APPB2007b}.

The success of our approach has its origin in  
the application of the concept of 
Clifford algebra to nonrelativistic phase space.
Although many authors 
stressed in various contexts
the importance of Clifford algebras in physics
(see e.g. \cite{Hestenes,Pavsic}), applications of this concept 
to phase space
are fairly rare.
It is therefore appropriate to study 
the structure of the Clifford algebra of nonrelativistic phase space
in some detail. 
This is all the more important because our phase space approach
necessarily introduces a fundamental constant of dimension [momentum/position] which
- when Planck constant is taken into account - should set a natural mass scale.
Thus, the hope is that our Clifford algebra contains not only
the generators of the
relevant symmetries and
the algebraic counterparts of positions and momenta,  
but will also provide us with some ideas concerning the algebraic approach to the
concept of mass. With this in mind,
we shall first classify all elements of the Clifford algebra in question
according to their $U(1)\otimes SU(3)$ and $SU(2)$ properties. 
 Then, we will linearize the nonrelativistic relation between kinetic energy, mass and momentum 
 and identify those elements of Clifford
algebra which may be associated with the concept of lepton mass.
Finally, using the lepton-quark transformations introduced in \cite{APPB2007b},
we will transform these elements from the lepton to the quark sector.
In this way, we obtain three elements of the Clifford algebra, each 
corresponding to the mass of an individual (coloured) quark.
We will then show that, although each of these elements, when taken separately, is not
invariant under ordinary rotation,
 the three elements add up to
give a rotationally-invariant total quark mass term.

We treat our
Clifford algebra of nonrelativistic phase space as a laboratory and testing
ground.
The real world is clearly much more complex than what this algebra suggests.
Yet, it should be pointed out that
our scheme treats the ordinary 
3D rotation in the only natural way, i.e. as {\em a pair of identical (same size and sense) 
rotations
in the momentum and position subspaces of the phase space}.
Consequently, all
  our conclusions regarding the behaviour under ordinary 3D rotations
  should be valid in general.

\section{Basic definitions, hypercharge and isospin}
We shall denote the basic elements of the relevant Clifford algebra by 
$A_k$ and $B_l$,
with $A_k~(B_l)$ associated with nonrelativistic momentum (position),
and use the following explicit representation:
\begin{eqnarray}
A_k&=&\sigma_k\otimes \sigma_0 \otimes \sigma_1, \nonumber\\
B_k&=&\sigma_0\otimes \sigma_k \otimes \sigma_2, \nonumber\\
B\equiv iA_1A_2A_3B_1B_2B_3&=&\sigma_0\otimes \sigma_0 \otimes \sigma_3.
\end{eqnarray}
In order to analyze our Clifford algebra in terms of its
 $U(1)\otimes SU(3)$ properties, it is appropriate to
introduce combinations analogous to the standard annihilation and creation operators
$a_k$ and $a^{\dagger}_k$, i.e.:
\begin{eqnarray}
C_k&=&\frac{1}{\sqrt{2}}(B_k+iA_k),\nonumber\\
C_k^{\dagger}&=&\frac{1}{\sqrt{2}}(B_k-iA_k).
\end{eqnarray}
The anticommutation relations satisfied by elements $A_k$, $B_l$, $B$
translate then into
\begin{eqnarray}
\{B,C_k\}=\{B,C^{\dagger}_k\}=\{C_k,C_l\}=\{C^{\dagger}_k,C^{\dagger}_l\}&=&0,\nonumber\\
{}\{C_k,C^{\dagger}_l\}&=&2\delta_{kl}.
\end{eqnarray}

The total of 64 elements of Clifford algebra 
may be grouped into four
sets of 16 elements each.
The first two sets are composed of linear combinations of the 
products of an even number of $A_k$, $B_l$, while
the latter two sets are built of linear combinations of 
the products of an odd number of $A_k$, $B_l$.
Before we proceed with the full presentation of all elements of the Clifford
algebra, we need to introduce two important even elements: 
the hypercharge $Y$ and the third component of the weak
isospin $I_3$.

\subsection{Hypercharge}
In line with \cite{APPB2007,Zen2008}, the hypercharge is defined as
\begin{equation}
Y=\frac{1}{3}\,{\cal{Y}},
\end{equation} 
where
(summation convention over repeated indices implied):
\begin{equation}
{\cal{Y}}=\sum_k{\cal{Y}}_k=-\frac{1}{2}[A_k,B_k]B=-\frac{1}{2}[C_k,C_k^{\dagger}]B.
\end{equation}
The $4 \times 4$ analogs of ${\cal{Y}}$ 
and ${\cal{Y}}_k$ will be denoted by $y$ and $y_k$:
\begin{equation}
\label{defY}
{\cal{Y}}= y \otimes \sigma_0=\sum_k{\cal{Y}}_k=\sum_{k=1}^3y_k\otimes \sigma_0=
\sigma_k\otimes\sigma_k\otimes\sigma_0.
\end{equation}
One finds that $y$ and $y_k$'s satisfy the following equations:
\begin{eqnarray}
y_1^2=y_2^2=y_3^2&=&+1,\nonumber\\
y_iy_j&=&-y_k~~~~(i \ne j \ne k \ne i),\nonumber\\
y_1y_2+y_2y_3+y_3y_1&=& -y,\nonumber\\
y_1y_2y_3&=&-1,\nonumber\\
\label{equationforY}
y ^2 + 2\, y - 3 &=& 0.
\end{eqnarray}
Matrices $y_k,y$  commute among themselves: 
$[y_k,y_l]=[y_k,y]=0$ for
any $k,l$, and, consequently, they may be simultaneously diagonalized.
Analogous statements hold for ${\cal{Y}}$ and ${\cal{Y}}_k$.
From the last equation in (\ref{equationforY}) it follows that 
the eigenvalues of $y $ are $+1$ (which is triple degenerate) and $-3$. 
The three ways of building the eigenvalue $y=+1$ out of 
the eigenvalues $\pm 1$ of $y_1$, $y_2$, and $y_3$
are identified with the colour degree of freedom.
The corresponding eigenvalues of $Y$ are $+\frac{1}{3}$ and $-1$,
as appropriate for the description of coloured quarks and leptons.
For more details, see \cite{APPB2007}-\cite{APPB2007b}.

\subsection{Weak isospin}
The weak isospin $I_3$ is related to the 7th anticommuting element of the algebra:
\begin{equation}
I_3=\frac{1}{2}B.
\end{equation}
Furthermore
(using the convention that underlined repeated indices are not summed over),
we define the following products of $B$, ${\cal{Y}}$, and ${\cal{Y}}_k$:
\begin{eqnarray}
R_k&=&{\cal{Y}}_kB=-\frac{1}{2}[C_{\underline{k}},C^{\dagger}_{\underline{k}}],\nonumber\\
R&=&{\cal{Y}}B=-\frac{1}{2}[C_{{k}},C^{\dagger}_{{k}}].
\end{eqnarray}
As is easily checked, all elements introduced so far, i.e. ${\cal{Y}}$, ${\cal{Y}}_k$,
$B$, $R$, and $R_l$, commute with each other.

\subsection{Projection operators}
In the following, we will need projection operators for the subspaces of definite $I_3$ and
$Y$.
For the $I_3=\pm\frac{1}{2}$ isospin subspaces, they are
given by:
\begin{equation}
I_{\pm \frac{1}{2}}=\frac{1}{2}\pm I_3.
\end{equation}
For the subspaces of hypercharge $Y=-1,+\frac{1}{3}$ 
the projection operators are
\begin{eqnarray}
Y_{-1}&=&\frac{1-{\cal{Y}}}{4},\nonumber\\
Y_{+\frac{1}{3}}&=&\frac{3+{\cal{Y}}}{4}.
\end{eqnarray}
As discussed in \cite{APPB2007}-\cite{APPB2007b}, the colour subspace \# $k$ is characterized by the set of
eigenvalues ($y_k=-1$, $y_{l_i}=+1$ for $l_{1,2} \ne k$).
The relevant projection operator is then
\begin{eqnarray}
Y_{+\frac{1}{3},k}&=&\frac{1}{4}(1+{\cal{Y}}-2{\cal{Y}}_k)=
\frac{3+{\cal{Y}}}{4}\cdot\frac{1-{\cal{Y}}_k}{4},
\end{eqnarray}
and it obviously satisfies:
\begin{equation}
\sum_kY_{+\frac{1}{3},k}=Y_{+\frac{1}{3}}.
\end{equation}
In the subsequent sections, the products of projection operators in the $I_3=\pm 1/2$ and
$Y=-1,+1/3$ subspaces will often occur.
Consequently, it is appropriate to introduce the following
 compact notation:
\begin{eqnarray}
Y^{\pm}_{-1}&=&I_{\pm\frac{1}{2}}Y_{-1},\nonumber\\
Y^{\pm}_{+\frac{1}{3}}&=&I_{\pm\frac{1}{2}}Y_{+\frac{1}{3}},\nonumber\\
\label{projectors}
Y^{\pm}_{+\frac{1}{3},k}&=&I_{\pm\frac{1}{2}}Y_{+\frac{1}{3},k}.
\end{eqnarray}

\section{Even elements of Clifford algebra}

The fifteen generators of $SU(4)$ are represented 
via the commutators among $A_k$, $B_l$. In terms of $C_k$, $C^{\dagger}_l$,
the relevant shift operators are expressed 
 as follows:
\begin{eqnarray}
H_{kl}&=&-\frac{1}{4}[C_k,C^{\dagger}_l]=(H^{}_{lk})^{\dagger},\nonumber\\
\label{Hm0H0mdef}
H_{m0}&=&-\frac{1}{8}\epsilon_{mkl}[C_k,C_l]=-\frac{1}{4}\epsilon_{mkl}C_kC_l,
\nonumber\\
H_{0m}&=&+\frac{1}{8}\epsilon_{mkl}[C^{\dagger}_k,C^{\dagger}_l]=
+\frac{1}{4}\epsilon_{mkl}C^{\dagger}_kC^{\dagger}_l.
\end{eqnarray}
The set of $H_{kl}$'s contains the $U(1)$ generator $H_{kk}$ and
the eight traceless SU(3) shift operators $H_{kl}-\frac{1}{3}\delta_{kl}H_{mm}$.
 Elements $H_{m0}$ and $H_{0m}$ constitute the ``genuine" $SU(4)$ shift
 operators.

Since elements $H_{kl}$, $H_{m0}$, and $H_{0m}$ commute with $I_3$, 
it is natural to introduce 
their projections onto the $I_3=\pm\frac{1}{2}$
subspaces:
\begin{eqnarray}
H^{\pm}_{nk}&=&H^{}_{nk}I_{\pm\frac{1}{2}},\nonumber\\
H^{\pm}_{n0}&=&H^{}_{n0}I_{\pm\frac{1}{2}},\nonumber\\
H^{\pm}_{0n}&=&H^{}_{0n}I_{\pm\frac{1}{2}},\nonumber\\
I_{\pm \frac{1}{2}}&=&1\cdot I_{\pm \frac{1}{2}}.
\end{eqnarray}
Thus, the 32 even elements are divided into two commuting sets composed of 16 elements
each, corresponding to sectors of given $I_3=\pm \frac{1}{2}$.

\subsection{$U(1) \otimes SU(3)$ generators}
\subsubsection{The $U(1) \otimes SU(3)$ structure}
The standard $SU(3)$ generators $F_b$'s ($b=1,...8$) are built from $H_{nk}$ as in Eq. (68) of
\cite{APPB2007}, with explicit form of $H_{nk}$'s given in Appendix A.
They satisfy
\begin{equation}
\label{FFcomrule}
[F^{\tau}_a,F^{\tau}_b]=2if_{abc}F^{\tau}_c,
\end{equation}
where $f_{abc}$ are SU(3) structure constants (see \cite{APPB2007}),
with $F^{\pm}_a=F_aI_{\pm \frac{1}{2}}$ being the projections of $F_a$ onto 
the $I_3=\pm \frac{1}{2}$ subspaces, 
and the two sets operating in disjoint subspaces:
\begin{equation}
F^{\pm}_b F^{\mp}_c=0. 
\end{equation} 
Since all $F_b$'s commute with $Y$, it is also natural to consider
the projections of $F^{\tau}_b$ onto the triplet and singlet subspaces. With
\begin{equation}
\label{twosidedsinglet}
Y_{-1}F^{\,\tau}_b=F^{\,\tau}_bY_{-1}=0,
\end{equation}
it follows that $F^{\,\tau}_b$'s  
are equal to their projections onto the triplet subspace:
\begin{eqnarray}
F^{\,\tau}_b&=&Y_{+\frac{1}{3}}F^{\,\tau}_bY_{+\frac{1}{3}}.
\end{eqnarray}

Together, the eight generators $F^{}_b$ of $SU(3)$ 
and the six ``genuine" $SU(4)$ shift operators $H_{m0}$, $H_{0m}$
 make fourteen generators.
The fifteenth generator of $SU(4)$ is
proportional to the $U(1)$ generator $R^{}$, and in the same
normalization 
it is:
\begin{equation}
F^{}_{15}\equiv \frac{1}{\sqrt{6}}\,R^{}=
\sqrt{6}\,YI_3,
\end{equation}
with 
 $R=2H_{kk}=\sigma_k\otimes\sigma_k\otimes\sigma_3$,
and its projections $R^{\pm}=RI_{\pm \frac{1}{2}}$
commuting with $F_b^{\tau}$:
\begin{equation}
[R^{\pm},F_b^{\tau}]=0.
\end{equation}
The four projection operators from the first two lines of Eq. (\ref{projectors})
are constructed from elements 
$R^{\pm}$ and $I_{\pm \frac{1}{2}}$.

If $X$ may be represented as a product of a certain definite
number of elements $C_k$ and $C^{\dagger}_l$, its $U(1)$
properties are specified by
 $\nu (X)$ defined as
\begin{equation}
[R,X]=+2\nu (X) X.
\end{equation}
The value of $\nu (X)$ gives 
the number of $C^{\dagger}_k$'s minus the number of $C_l$'s present
 in  the product in question.
 Thus, $\nu (F_b)=0$.\\

\subsubsection{Transformations of $C_k$ and $C^{\dagger}_k$}
Shift operators $H_{kl}$ act on $C_n$ and $C^{\dagger}_n$ as
follows (for any $k$, $l$, $n$):
\begin{eqnarray}
{}[H_{kl},C_n]&=&-\delta_{ln}C_k,\nonumber\\
\label{HwithCandCdagger}
{}[H_{kl},C^{\dagger}_n]&=&+\delta_{kn}C^{\dagger}_l.
\end{eqnarray}
In particular, under SU(3) transformations, elements $C_k$ transform as a triplet, 
while $C^{\dagger}_k$ as an antitriplet:
\begin{eqnarray}
[F_a,C_k]&=&\lambda_{akl}C_l,\nonumber\\{}
\label{FCkcompact}
[F_a,C^{\dagger}_k]&=&-\lambda^*_{akl}C^{\dagger}_l,
\end{eqnarray}
where $\lambda_{akl}$ may be read off from:
\begin{eqnarray}
[F_1,C_2]=[F_3,C_1]=[F_4,C_3]
=\sqrt{3}[F_8,C_1]&=&-C_1,\nonumber\\
{}[F_1,C_1]=-[F_3,C_2]=[F_6,C_3]
=\sqrt{3} [F_8,C_2]&=&-C_2,\nonumber\\
{}[F_4,C_1]=[F_6,C_2]=
-\frac{\sqrt{3}}{2}[F_8,C_3]&=&-C_3,\nonumber\\
{}[F_2,C_2]=[F_5,C_3]&=&-iC_1,\nonumber\\
{}[F_2,C_1]=-[F_7,C_3]&=&+iC_2,\nonumber\\
\label{FCk}
{}[F_5,C_1]=[F_7,C_2]&=&+iC_3,
\end{eqnarray}
(the remaining commutators are zero).
For $C_k^{\dagger}$ (antitriplet) the relevant commutation relations are obtained by taking
the hermitian conjugate of the above. 
Under $U(1)$ we have:
\begin{eqnarray}
[R^{},C_k]&=&-2C_k,\nonumber\\
\label{RCk}
{}[R^{},C^{\dagger}_k]&=&+2C^{\dagger}_k,
\end{eqnarray}
i.e. $\nu(C_k)=-\nu(C^{\dagger}_k)=-1$.
Under ordinary reflections ${\cal{P}}=\exp (-i\frac{\pi}{2}R)=-iB$, element $C_k$ 
($C_k^{\dagger}$) changes sign:
\begin{equation}
{\cal{P}}C_k{\cal{P}}^{-1}=-C_k.
\end{equation}

\subsection{Genuine generators of SU(4)}
\label{SU4section}
The six additional shift operators $H_{m0}$ and $H_{0m}$
satisfy
\begin{equation}
H_{0m}=(H_{m0})^{\dagger}.
\end{equation}
and constitute ``genuine" SU(4) operators.
Their explicit forms are
 given in Appendix A.
 
Shift operators 
$H^{\tau}_{m0}$ and $H^{\tau}_{0m}$
act between 
subspaces with different eigenvalues of $Y$:
\begin{eqnarray}
H^{+}_{m0}&=&Y^+_{-1}
H_{m0}Y^+_{+\frac{1}{3}},
\nonumber\\
H^{-}_{m0}&=&Y^-_{+\frac{1}{3}}
H_{m0}Y^-_{-1},
\nonumber\\
\label{H0m+-}
H^{+}_{0m}&=&Y^+_{+\frac{1}{3}}
H_{0m}Y^+_{-1},
\nonumber\\
H^{-}_{0m}&=&Y^-_{-1}
H_{0m}Y^-_{+\frac{1}{3}},
\end{eqnarray}
i.e. 
they connect the triplet and singlet 
$SU(3)$ subspaces with each other, as also indicated by subscripts
``$m0$" and ``$0m$" of our notation. 
The six ``genuine" hermitean 
generators of $SU(4)$ are built from $H^{\tau}_{m0}$ and $H^{\tau}_{0m}$ as:
\begin{eqnarray}
F^{\tau}_{+n}&=&H_{n0}^{\tau}+H^{\tau}_{0n},
\nonumber\\
F^{\tau}_{-n}&=&i(H_{n0}^{\tau}-H^{\tau}_{0n}). 
\end{eqnarray}
Elements $F^{\tau}_{-n}$ describe simultaneous 
 rotations in ${\bf x}$ and ${\bf p}$
spaces in mutually opposite senses 
(counterparts to ordinary simultaneous rotations in likewise
senses).

\subsubsection{Transformation properties of $H_{m0}$ and $H_{0m}$}

The $U(1) \otimes SU(3) $ transformation properties of 
$H_{m0}$ and $H_{0m}$ are (for any $k,l,m$):
\begin{eqnarray}
[H^{\tau}_{kl},H^{\tau}_{m0}]&=&+\delta_{mk}H^{\tau}_{l0}-
\delta_{kl}H^{\tau}_{m0},\nonumber\\
\label{Hm0H0mSU3prop}
{}[H^{\tau}_{kl},H^{\tau}_{0m}]&=&-\delta_{ml}H^{\tau}_{0k}+
\delta_{kl}H^{\tau}_{0m}.
\end{eqnarray}
From the point of view of $SU(3)$ 
(traceless generators, i.e. either $k \ne l$, or appropriate linear combinations
of terms with $k =l$), the second term on the r.h.s. above does not contribute.
Thus, the above equation shows that 
the $SU(3)$ transformation properties of $H_{m0}^{\tau}$ 
coincide with those of $C^{\dagger}_m$
in Eq. (\ref{HwithCandCdagger}), i.e. with antitriplet, 
while the $H_{0m}$'s transform like $C_m$, 
i.e. an $SU(3)$
triplet.
In other words, we have:
\begin{eqnarray}
[F^{\tau}_a,H^{\tau}_{0k}]&=&\lambda_{akm}H^{\tau}_{0m},\nonumber\\{}
[F^{\tau}_a,H^{\tau}_{k0}]&=&-\lambda^*_{akm}H^{\tau}_{m0}.
\end{eqnarray}
From Eq. (\ref{Hm0H0mSU3prop}) we further obtain that 
elements $H^{\tau}_{m0}$ and $H^{\tau}_{0m}$
transform under $U(1)$ like
\begin{eqnarray}
[R^{\tau},H^{\tau}_{m0}]&=&-4H^{\tau}_{m0},\nonumber\\
{}[R^{\tau},H^{\tau}_{0m}]&=&+4H^{\tau}_{0m},
\end{eqnarray}
i.e. $H^{\tau}_{m0}$ 
transform 
like a simple product of two $C_k$'s (and not like a single $C^{\dagger}_m$),
with $R$ eigenvalues of $C_k$'s simply added, while 
$H^{\tau}_{0m}$ transform like a product of two $C^{\dagger}_k$'s
(see Eqs (\ref{[H0m,Cn]})). Obviously
\begin{equation}
\nu(H^{\tau}_{0m})=-\nu(H^{\tau}_{m0})=+2.
\end{equation}

For completeness, we also give the anticommutators (for any $k,l,m$):
\begin{eqnarray}
\{H^{\pm}_{kl},H^{\pm}_{m0}\}&=&\mp \delta_{km}H^{\pm}_{l0},\nonumber\\
\{H^{\pm}_{kl},H^{\pm}_{0m}\}&=&\mp \delta_{lm}H^{\pm}_{0k},
\end{eqnarray}
and the products of genuine  $SU(4)$ shift operators themselves:
\begin{eqnarray}
H^{\tau}_{m0}H^{\tau}_{n0}&=&H^{\tau}_{0m}H^{\tau}_{0n}=0,\nonumber\\
H_{0n}H_{m0}&=&\frac{1}{4}\,\delta_{nm}\,(1+R)-H^+_{nm},\nonumber\\
H_{m0}H_{0n}&=&\frac{1}{4}\,\delta_{nm}\,(1-R)+H^-_{nm}.
\end{eqnarray}
From the latter formulas one gets
\begin{eqnarray}
[H^{\tau}_{0n},H^{\tau}_{m0}]&=&\frac{1}{2}\delta_{nm}R^{\tau}-H^{\tau}_{nm},\nonumber\\{}
\{H^{\tau}_{0n},H^{\tau}_{m0}\}&=&
\frac{1}{2}\,\delta_{nm}\,I_{\tau\frac{1}{2}}-\tau H^{\tau}_{nm}.
\end{eqnarray}

\begin{table}[t]
\caption{$U(1)\otimes SU(3)$ classification of 32 even elements
of Clifford algebra. In the four rightmost columns the relevant 
left and right eigenvalues of $Y$ and $Q$ are given.}
\label{tableFH}
\begin{center}
\begin{math}
\begin{array}{ccccccc}
\multicolumn{2}{c}{{\rm Sector}~I_{3}=+\frac{1}{2}}&&&&&\\
&&&&&&\\
U(1)&SU(3)&~~~~&~~~Y_l~~&~~Y_r~~&~~Q_l~~&~~Q_r~ \\
\hline 
-2&\bar{3}&H^{+}_{m0} &-1&+\frac{1}{3}&0&+2/3\rule{0mm}{5mm}\\
+2&3&H^{+}_{0m} &+\frac{1}{3}&-1&+2/3&0\rule{0mm}{5mm}\\
0&8&F^{+}_b &+\frac{1}{3}&+\frac{1}{3}&+2/3&+2/3\rule{0mm}{5mm}\\
\hline
0&1&Y^+_{-1}&-1&-1&0&0\rule{0mm}{5mm}\\
0&1&Y^+_{+\frac{1}{3}}&+\frac{1}{3}&+\frac{1}{3}&+2/3&+2/3\rule{0mm}{5mm}\\
\hline
\hline
&&&&&&\\
\multicolumn{2}{c}{{\rm Sector}~I_{3}=-\frac{1}{2}}&&&&&\\
&&&&&&\\
U(1)&SU(3)&~~~~&~~~Y_l~~&~~Y_r~~&~~Q_l~~&~~Q_r~ \\
\hline
-2&\bar{3}&H^{-}_{m0} &+\frac{1}{3}&-1&-1/3&-1\rule{0mm}{5mm}\\
+2&3&H^{-}_{0m} &-1&+\frac{1}{3}&-1&-1/3\rule{0mm}{5mm}\\
0&8&F^{-}_b &+\frac{1}{3}&+\frac{1}{3}&-1/3&-1/3\rule{0mm}{5mm}\\
\hline
0&1&Y^-_{-1}&-1&-1&-1&-1\rule{0mm}{5mm}\\
0&1&Y^-_{+\frac{1}{3}}&+\frac{1}{3}&+\frac{1}{3}&-1/3&-1/3\rule{0mm}{5mm}\\
\hline
\hline
\end{array}
\end{math}
\end{center}
\end{table}

\subsubsection{Transformations of $C_k$ and $C_k^{\dagger}$}
Under the action of $H_{m0}$ and $H_{0m}$, matrices $C_n$
and $C^{\dagger}_n$ transform as follows:
\begin{eqnarray}
[H_{m0},C_n]&=&0,\nonumber\\{}
[H_{0m},C_n]&=&-\epsilon_{mnj}C^{\dagger}_j,\nonumber\\{}
[H_{0m},C^{\dagger}_n]&=&0,\nonumber\\{}
\label{[H0m,Cn]}
[H_{m0},C^{\dagger}_n]&=&+\epsilon_{mnj}C_j.
\end{eqnarray}
This translates into:
\begin{eqnarray}
{}[F^{}_{+n},C_k]&=&-\,\epsilon_{nkl}C^{\dagger}_l,\nonumber\\
\label{F+n}
{}[F^{}_{+n},C^{\dagger}_k]&=&+\,\epsilon_{nkl}C_l,
\end{eqnarray}
and
\begin{eqnarray}
{}[F^{}_{-k},C_l]&=&+i\,\epsilon_{klm}C^{\dagger}_m,\nonumber\\
\label{F-n}
{}[F^{}_{-k},C^{\dagger}_l]&=&+i\,\epsilon_{klm}C_m.
\end{eqnarray}
To summarize, the 32 even elements of the Clifford algebra are composed of the
unit element and the 15 generators of SU(4), 
with each of these 16 elements multiplied by $I_{\pm\frac{1}{2}}$.
These two sets commute with each other.
Under the $SU(3)$ transformations each set decomposes  into 
two singlets (i.e. projection operators), an octet,
 a triplet and an antitriplet. 
 All elements stay invariant under ordinary reflection.
 The full decomposition is given in Table \ref{tableFH}.
 The first two columns of the table describe transformation properties under $U(1)$ 
 (the value of $\nu$) and $SU(3)$ (representation). The four rightmost columns specify
 the left and right eigenvalues of $Y$ and $Q$.

\section{Odd elements of Clifford algebra}
The even elements of Clifford algebra are diagonal in $I_3$. In order to discuss
the odd elements, we define weak isospin raising and lowering operators $I_+$ and $I_-$:
\begin{equation}
\label{odd}
I_+=\sigma_0\otimes \sigma_0 \otimes \frac{\sigma_1+i\sigma_2}{\sqrt{2}}
\propto A_1A_2A_3+iB_1B_2B_3,
\end{equation}
with $I_-=I_+^{\dagger}$.
With elements $I_{\pm}$ being odd, the odd elements of Clifford algebra, i.e.
sets 3) and 4) are then obtained via multiplication by 
$I_{\pm}$ of the (even) elements of 
the first two sets.  Obviously, the odd elements are off-diagonal in $I_3$ and change sign under ordinary
reflection.
All odd elements may be obtained from products of an odd
number (one or three) of $C_k$'s  and $C^{\dagger}_l$,
with these products multiplied from left and right 
by the (even) projection operators
 corresponding to subspaces of definite $Y$ and $I_3$.

\subsection{SU(3) triplets and antitriplets}
We now project $C_k$ (either from the left or from the right) 
onto subspaces of definite $Y$ and $I_3$, and
define $W_k,V_k,U_k$  as follows:
\begin{eqnarray}
W_k&=&i\,Y^+_{-1}C_k=i\,C_kY^-_{+\frac{1}{3}},
\nonumber\\
V_k&=&i\,Y^+_{+\frac{1}{3}}C_k=i\,C_kY^-_{-1},
\nonumber\\
U_k&=&i\,Y^-_{+\frac{1}{3}}C_k=
i\,C_kY^+_{+\frac{1}{3}},
\nonumber\\
\label{WVUprojections}
0&=&Y^-_{-1}C_k=C_kY^+_{-1}.
\end{eqnarray}
They satisfy
\begin{equation}
\label{W+V+U}
W_k+V_k+U_k=iC_k.
\end{equation}
Explicit expressions for $U_k,V_k,W_k$ are given in Appendix A.

Since $F^{}_a$ commute with $I_3$ and $Y$, it follows that
$U_k$, $V_k$, and $W_k$ transform under $SU(3)$ just like $C_k$, i.e. they
are $SU(3)$ triplets. Thus, we have:
\begin{eqnarray}
[F_a,V_k]&=&F^+_aV_k=\lambda_{akl}V_l,\nonumber\\{}
[F_a,W_k]&=&-W_kF^-_a=\lambda_{akl}W_l,\nonumber\\{}
[F_a,U_k]&=&F^{-}_aU_k-U_kF^{+}_a=\lambda_{akl}U_l.
\end{eqnarray}
The hermitean conjugates $U^{\dagger}_k$, $V^{\dagger}_k$, 
and $W^{\dagger}_k$ transform like $C_k^{\dagger}$ and are antitriplets.
Similarly, since $R$ commutes with $Y$ and $I_3$,
it follows that
$U_k$, $V_k$, and $W_k$ transform under $U(1)$ just like $C_k$:
\begin{eqnarray}
[R,U_k]&=&-2U_k,\nonumber\\
{}[R,V_k]&=&-2V_k,\nonumber\\
{}[R,W_k]&=&-2W_k,
\end{eqnarray}
i.e. the values of $\nu$ are still well defined:
\begin{equation}
\nu(U_k)=\nu(V_k)=\nu(W_k)=-1.
\end{equation}

Multiplication rules for elements $U_k$, $V_l$, $W_m$ and their hermitean conjugates 
are given in Appendix B.
When expressed in terms of $U_k$, $V_l$ and $W_m$, 
the $U(1) \otimes SU(3)$ generators in $I_3=\pm \frac{1}{2}$ subspaces are:
\begin{eqnarray}
H^+_{kl}&=&-\frac{1}{4}
(V_kV_l^{\dagger}+W_kW_l^{\dagger}-U_l^{\dagger}U_k),\nonumber\\
H^-_{kl}&=&+\frac{1}{4}
(V_l^{\dagger}V_k+W_l^{\dagger}W_k-U_kU_l^{\dagger}).
\end{eqnarray}
For the genuine $SU(4)$ shift operators, we have:
\begin{eqnarray}
H^{+}_{m0}&=&
=+\frac{1}{4}\epsilon_{mkl}W_kU_l,\nonumber\\
H^{-}_{m0}&=&
=+\frac{1}{4}\epsilon_{mkl}U_kV_l,\nonumber\\
H^{+}_{0m}&=&
=-\frac{1}{4}\epsilon_{mkl}U^{\dagger}_kW^{\dagger}_l,\nonumber\\
H^{-}_{0m}&=&
=-\frac{1}{4}\epsilon_{mkl}V^{\dagger}_kU^{\dagger}_l.
\end{eqnarray}

\subsection{SU(3) singlets}
The only nonzero products that one can form from $C_k$'s are
$C_kC_l$ (i.e. $H^{}_{m0}$) and the totally
antisymmetric product
$C_1C_2C_3$. We now define:
\begin{eqnarray}
\label{Easmixedprod}
\epsilon_{mkn}G_0&=& \frac{1}{2}C_{m}C_kC_{n}=\frac{1}{8}\{[C_m,C_k],C_n\},
\end{eqnarray}
with the mixed product 
$ \{[C_m,C_k],C_n\}$ 
satisfying
\begin{eqnarray}
\label{CCC1}
\{[C_k,C_n],C_m\}=&\{[C_m,C_k],C_n\}&
=\{[C_n,C_m],C_k\}.
\end{eqnarray}
The explicit form of $G_0$ is given in Appendix A.

Using Eq (\ref{Hm0H0mdef}) we may rewrite $G_0$ also as
\begin{eqnarray}
G_0&=&-\frac{1}{2}\{H_{\underline{k}0},C_{\underline{k}}\}
=\frac{1}{16}\epsilon_{mn\underline{k}}\{[C_m,C_n],C_{\underline{k}}\}.
\label{G0Hk0Ck}
\end{eqnarray}
Element $G_0$ is diagonal in $Y$:
\begin{eqnarray}
G_0&=&
Y^+_{-1}G_0=G_0Y^-_{-1},\nonumber\\
0&=&
Y^-_{-1}G_0=G_0Y^+_{-1}.
\end{eqnarray}
Thus, $G_0$ (and $G^{\dagger}_0$) require ${{Y}}=-1$ and correspond to
leptons.
With $Y^{\pm}_{-1}$ being projection operators onto the SU(3) 
singlet subspace, it is obvious from (\ref{twosidedsinglet})
that $G_0$ is a $SU(3)$ singlet :
\begin{equation}
\label{E0singlet}
[F^{}_b,G_0]=0.
\end{equation}
This is also seen from Eq. (\ref{G0Hk0Ck}) 
which (when summed over $k$) contains a trace of the product of a triplet $C_k$
and an antitriplet $H^{}_{k0}$.
Element $G_0$ transforms under $U(1)$  as (e.g. using
Eq. (\ref{RCk})):
\begin{equation}
[R,G_0]=-6G_0,
\end{equation}
i.e. $\nu(G_0)=-3$.

In terms of $U_k$, $V_l$, and $W_m$, we have:
\begin{eqnarray}
G_0&=&+\frac{i}{12}\,\epsilon_{mkn}W_mU_kV_n,\nonumber\\
\label{E0}
G^{\dagger}_0&=&
+\frac{i}{12}\,\epsilon_{mkn}V^{\dagger}_mU^{\dagger}_kW^{\dagger}_n,
\end{eqnarray} 
i.e., $G_0,G^{\dagger}_0$ are proportional to weak isospin raising 
and lowering operators
in lepton subspace (consult the left and right eigenvalues of $I_3$ for $W_k$ and
$V_l$, Eq. (\ref{WVUprojections})).

\subsection{SU(3) sextets and antisextets}
In analogy to Eq. (\ref{G0Hk0Ck}), we now form elements $G_{\{kl\}}$
defined as
(for any $k,l$):
 \begin{eqnarray}
 \label{generalGkl}
 G_{\{kl\}}
 &=&\frac{1}{4}\left(\{H_{0k},C_l\}+\{H_{0l},C_k\}\right).
 \end{eqnarray}
Element $G_{\{kl\}}$ is built as a symmetric combination of a product of two triplets:
$C_l$ and $H_{0k}$, and, consequently, it is a sextet.
Its h.c. element  $G^{\dagger}_{\{kl\}}$ is an antisextet. Their explicit forms are
given in Appendix A.

Under $U(1) \otimes SU(3)$ the sextet transforms as:
\begin{equation}
[H_{kl},G_{\{mn\}}]=
-\delta_{lm}G_{\{kn\}}-\delta_{ln}G_{\{km\}}+\delta_{kl}G_{\{mn\}}.
\end{equation}
For the $SU(3)$ (i.e. traceless) generators the last term on the r.h.s. above does
not contribute. It contributes only when one evaluates the 
commutator of the sum $H_{kk}$ with $G_{\{mn\}}$, which  leads to
the following behaviour of $G_{\{mn\}}$ under $U(1)$:
\begin{equation}
[R,G_{\{mn\}}]=+2G_{\{mn\}},
\end{equation}
i.e. $\nu (G_{\{mn\}})=+1$.

Furthermore, for any $k$, $l$ one has:
\begin{eqnarray}
G_{\{kl\}}&=&Y^+_{+\frac{1}{3}}G_{\{kl\}}=G_{\{kl\}}Y^-_{+\frac{1}{3}},\nonumber\\
0&=&Y^-_{+\frac{1}{3}}G_{\{kl\}}=G_{\{kl\}}Y^+_{+\frac{1}{3}},
\end{eqnarray}
and similarly for $G^{\dagger}_{\{kl\}}$.
Thus, $G_{\{kl\}}$ is equal to its projection onto the $Y=+\frac{1}{3}$ (i.e. quark)
subspace.
\begin{table}[t]
\caption{$U(1)\otimes SU(3)$ classification of 32 odd elements
of Clifford algebra. 
In the four rightmost columns the relevant left and right eigenvalues of 
$Y$ and $Q$ are given.}
\label{tableUVW}
\begin{center}
\begin{math}
\begin{array}{ccccccc}
\multicolumn{3}{c}{{\rm Sector}~ I_{3,l}=+\frac{1}{2}, ~I_{3,r}=-\frac{1}{2}}&&&&\\
&&&&&&\\
U(1)&SU(3)&~~~~&~~~Y_l~~&~~Y_r~~&~~~Q_l~~&~~Q_r~ \\
\hline 
+1&\bar{3}&U^{\dagger}_k &+\frac{1}{3}&+\frac{1}{3}&+2/3&-1/3\rule{0mm}{5mm}\\
-1&3&V_k &+\frac{1}{3}&-1&+2/3&-1\rule{0mm}{5mm}\\
-1&3&W_k &-1&+\frac{1}{3}&0&-1/3\rule{0mm}{5mm}\\
\hline
+1&6&G_{\{kl\}}&+\frac{1}{3}&+\frac{1}{3}&+2/3&-1/3\rule{0mm}{5mm}\\
-3&1&G_0&-1&-1&0&-1\rule{0mm}{5mm}\\
\hline
\hline
&&&&&&\\
\multicolumn{3}{c}{{\rm Sector}~ I_{3,l}=-\frac{1}{2},~I_{3,r}=+\frac{1}{2}} &&&&\\
&&&&&&\\
U(1)&SU(3)&~~~~&~~~Y_l~~&~~Y_r~~&~~~Q_l~~&~~Q_r~ \\
\hline 
-1&3&U_k &+\frac{1}{3}&+\frac{1}{3}&-1/3&+2/3\rule{0mm}{5mm}\\
+1&\bar{3}&V^{\dagger}_k &-1&+\frac{1}{3}&-1&+2/3\rule{0mm}{5mm}\\
+1&\bar{3}&W^{\dagger}_k &+\frac{1}{3}&-1&-1/3&0\rule{0mm}{5mm}\\
\hline
-1&\bar{6}&G^{\dagger}_{\{kl\}}&+\frac{1}{3}&+\frac{1}{3}&-1/3&+2/3\rule{0mm}{5mm}\\
+3&1&G^{\dagger}_0&-1&-1&-1&0\rule{0mm}{5mm}\\
\hline
\hline
\end{array}
\end{math}
\end{center}
\end{table}
 
When $k=l$ Eq. (\ref{generalGkl}) reduces to
 \begin{eqnarray}
 \label{Gnn}
 G_n\equiv G_{\{\underline{nn}\}}&=
 &+\frac{1}{2}\{H_{0\underline{n}},C_{\underline{n}}\}
 =\frac{1}{16}\epsilon_{mk\underline{n}}
 \{[C^{\dagger}_m,C^{\dagger}_k],C_{\underline{n}}\}, 
 \end{eqnarray}
 which constitutes a counterpart of Eq (\ref{G0Hk0Ck})
 with mixed product $\{[C^{\dagger}_m,C^{\dagger}_k],C_n\}$
 replacing $\{[C^{}_m,C^{}_k],C_n\}$ and
 satisfying
\begin{eqnarray}
\{[C^{\dagger}_m,C^{\dagger}_k],C_n\}&=\{[C_n,C^{\dagger}_m],C^{\dagger}_k\}
=&\{[C^{\dagger}_k,C_n],C^{\dagger}_m\}.
\end{eqnarray}

Explicitly, we have
\begin{eqnarray}
G_n&=&
Y_{+\frac{1}{3},n}I_{+},\nonumber\\
\label{En}
G^{\dagger}_n&=&
Y_{+\frac{1}{3},n}I_{-},
\end{eqnarray}
with $G_n$ ($G^{\dagger}_n$) containing projection operators onto colour subspace \# $n$.
These are essentially weak isospin raising and lowering operators for quarks of a
given colour.

In terms of $U_k$, $V_l$, $W_m$, the elements $G_{\{kl\}}$ and $G^{\dagger}_{\{kl\}}$ may be expressed in various ways 
(using relations (\ref{WWdagVdagV1})) with the simplest form being
\begin{eqnarray}
G_{\{kl\}}&=&-\frac{i}{8}\,U^{\dagger}_r\,
(\epsilon_{krs}U_l+\epsilon_{lrs}U_k)\,
U^{\dagger}_s,\nonumber\\
G_{\{kl\}}^{\dagger}&=&+\frac{i}{8}\,U^{}_s\,
(\epsilon_{krs}U^{\dagger}_l+\epsilon_{lrs}U^{\dagger}_k)\,
U^{}_r.
\end{eqnarray}

To summarize, under the SU(3) transformations the sixteen elements proportional to $\sigma_1+i\sigma_2$
decompose into: an antitriplet, two triplets, a sextet,  and a singlet.
Similar decomposition holds for the h.c. elements proportional to $\sigma_1-i\sigma_2$.
The full decomposition is given in Table \ref{tableUVW}.

\section{The concept of mass }
\subsection{From $U(1)\otimes SU(3)$ to $SO(3)$}
\label{masses}
The basic physical idea motivating our approach is that the
familiar macroscopic notions of space, time, etc. are emergent concepts, 
which do not exist at the ``true" quantum level in any form other than a
very rudimentary one. 
This idea was pursued in various contexts by
many. For instance, Penrose suggested spin 
 as a precursor of the concept of
direction in the 3D space \cite{Penroseangle}.  
Page and Wootters proposed that quantum
correlations could give rise to the macroscopic concept of time 
\cite{Wootterstime}.  
The general idea was succinctly expressed by Wheeler 
as ``Day One - quantum principle, Day Two - geometry"
\cite{WheelerDayOneDayTwo}.
It was also pointed out that causality and quantum prescriptions,
when combined,
suggest the existence (or emergence) of a preferred frame \cite{BellEberhard}
and absolute simultaneity.
I believe therefore that
mixing the ordinary 3D space with time, 
characteristic of the standard form of
special relativity, should not be used
as a starting point to seek the underlying "true" quantum level.
Instead, 
with quantum mechanics living in phase space, it is 
the mixing of the 3D space of positions with the 3D space of momenta which
should be more appropriate.
In line with these ideas, the concept of 3D
rotations in ordinary space - understood as same-size and -sense 
rotations in momentum and position
subspaces - is in our approach extended to 6D rotations in phase space.
This requires introduction of a new physical constant of dimension 
[momentum/position]. At the quantum level, with Planck constant at our disposal, the
mass scale is then set. 
This quantum level is therefore expected to contain not only spin,
 but also some quantum ideas about the concept of mass.
I believe therefore that a part of the problem of mass quantization  
should find its resolution at the level of 6D rotations in
the algebra underlying nonrelativistic phase space. In other words, I think that this algebra 
(or perhaps its appropriate
generalization) should lead to
 a joint quantum treatment of both masses and spins. 
  The approach of \cite{APPB2007}-\cite{Zen2006} and of this paper constitute but
a first step in that direction.
Speaking more generally, 
 the idea of ``space quantization" 
 (or rather "finding the underlying quantum-level precursors of 3D space")
 is naturally replaced 
 in the phase space approach by similarly understood
 ``phase-space quantization". 
 Thus, when ``space" is understood as ``phase-space", 
 the problem of ``space quantization" 
  and the problem of elementary particle mass spectrum (and of their quantum
  numbers) seem closely related.
  I am tempted to think of them as of 
  conceptually one and the same problem.

While the main idea of \cite{APPB2007}-\cite{APPB2007b} is concerned with rotations in
phase space, the treatment of reflections, although explicit, is extremely
simplified when compared to what the SM tells us about the quantum world of elementary particles. 
Yet, 
reflection at
the quantum level has to be properly treated if
the space (phase space) points are to emerge with their real-world properties.
This puts a question mark with respect to the introduction of the gauge principle into our
present scheme. 
One could perhaps satisfy oneself with a ``mixed" (i.e. classical-quantum) 
level of theory 
and introduce the gauge structure at least for the $U(1) \otimes SU(3)$
part. We should be then back at the field-theoretical SM level of description. 
 I think, however, that such a procedure
will be fully legitimate only when 
a better understanding of how to generate space (phase space) points is reached.
Consequently, in this paper we do not address the issue of gauge invariance.
The latter should appear in its full form only on "Day Three".

In the Standard Model the mass-generating prescription (i.e. Higgs mechanism)
 lies outside of the strict gauge 
structure of the theory. 
Gauge interactions tell us nothing about the masses of the fundamental particles. 
In fact, these masses have to be put into SM by hand.  
Thus, in the SM the problem of mass appears to be separate from gauge
interactions.
Consequently, in our approach we should also expect some separation between
the $U(1)\otimes SU(3)$ group structure (related to phase-space symmetries). and the mass generating mechanism.
The overall scheme is thus anticipated to contain the following two ingredients:\\
\phantom{xxx}1) The mass-independent part, i.e. the symmetry group 
$U(1) \otimes SU(3)$ (combined with $SU(2)$). It is related to the
nonrelativistic rotational (and reflectional) symmetry
generalized to $O(6)$, with the generalization
assumed valid at each point of our 3D world.
Consequently, it
should survive the emergence of space (and time) points
and constitutes a conjectured precursor of the SM gauge group.\\
\phantom{xxx}2) A separate mass-defining prescription.
This prescription must be tied to the standard concept of mass. This means it has to be
related to the choice 
of which three coordinates of the 6D phase space are to be considered as
the standard momentum coordinates, or, in other words, 
which
three of the eight $SU(3)$ generators constitute the ordinary
 rotations of our 3D macroscopic world.
At the quantum level, therefore, the 
relevant quantum-level
prescription for mass should be related to spin (and its quantization). 

The appearance of the standard
concept of mass is here linked with the reduction of 
the precursor of the unbroken
part of the SM gauge group to ordinary $O(3)$.  
This  may be considered strange.
Yet, it seems to follow naturally from the
assumptions of our approach, which look very appealing.
Thus, one feels forced to accept their consequences.
In summary, the problem of mass quantization is thought to 
 belong to the underlying quantum level. On the other hand,
the standard description
of electromagnetic and strong interactions 
 is expected to
 be the emergent one, with the underlying $U(1)\otimes SU(3)$ symmetry group
 conjectured to survive the procedure of mass quantization in the form of gauge group.
\subsection{Lepton mass term}
As argued above, we must tie our scheme to the standard concept of mass.
Now, lepton mass is certainly standard since it appears in a relation 
which connects lepton energy (kinetic energy in nonrelativistic case), 
momentum and mass.
Consequently, we must start from the lepton mass term.
In order to discuss it, we first note that in our Clifford
algebra
there are only four elements which are both diagonal in $Y$
and have $Y=-1$, as appropriate for the leptons. These are the 
(even) projection operators $Y^{\pm}_{-1}$ and the (odd)
elements $G_0$ and $G^{\dagger}_0$. The latter two terms are $SU(3)$ singlets
and $SO(3)$ scalars and, being odd like $A_k$ (which is related to momentum), 
they constitute natural candidates for
the elements related to lepton mass and kinetic energy.

As we show below, the nonrelativistic expression relating
kinetic energy, mass and momentum gets linearized in our algebra
by the consideration of the following linearized forms:
\begin{eqnarray}
{\cal{L}}_1&=&Y_{-1}(A_kp_k-m_1G_0+EG_0^{\dagger}),\nonumber\\
{\cal{L}}_2&=&(A_kp_k-m_2G_0+EG_0^{\dagger})Y_{-1}.
\end{eqnarray}
(Obviously, we could
consider similar forms with $G_0 \leftrightarrow G^{\dagger}_0$.)
The $Y_{-1}$ projection operator appears here because
our goal is the treatment of leptons.
The presence of $Y_{-1}$ (which commutes with ordinary 3D rotations) is irrelevant for the $G_0$ and $G_0^{\dagger}$ terms since
$Y_{-1}G_0=G_0$, $Y_{-1}G^{\dagger}_0=G^{\dagger}_0$, etc., but it affects the terms
containing $A_k$.

Elements $G_0$ and $G_0^{\dagger}$ are not invariant under reflection.
This is not a feature expected for mass (energy) terms.
Yet, a similar lack of reflection invariance was observed in a fully-fledged Galilean 
framework \cite{HK2003}.
This might be thought of as an indication that in the nonrelativistic Clifford algebra
approach
the treatment of reflection is 
 oversimplified, as argued in the previous subsection.
On the other hand, this persistent
appearance of the violation of parity invariance  
might also be considered an 
interesting feature of the nonrelativistic approach. In any case, 
whatever point of view  concerning the treatment of reflections one adopts, 
all our conclusions regarding
the rotational properties should stay unaffected.

Since $Y_{-1}A_kY_{-1}\propto Y_{-1}(C^{\dagger}_k-C_k)Y_{-1} = 0$ 
it follows that
\begin{eqnarray}
{\cal{L}}_1{\cal{L}}_2&=&
Y_{-1}(A_kA_np_kp_n-m_1EG_0G_0^{\dagger}-m_2EG^{\dagger}_0G_0)Y_{-1}
\nonumber\\
&=&Y_{-1}(A_kA_np_kp_n-2m_1E\,Y^+_{-1}
-2m_2E\,Y^-_{-1})Y_{-1}\nonumber\\
&=&Y^+_{-1}({\bf p}^2-2m_1E)+
Y^-_{-1}({\bf p}^2-2m_2E)
\end{eqnarray}
where we used Eq. (\ref{G0G0+}), i.e.
\begin{eqnarray}
(G_{0})^2=(G^{\dagger}_{0})^2&=&0,\nonumber\\
G_{0}G^{\dagger}_{0}&=&2Y^+_{-1},\nonumber\\
\label{E0E0dagger}
{}G^{\dagger}_{0}G_{0}&=&2Y^-_{-1}.
\end{eqnarray}
Thus, nonrelativistic expressions for kinetic energies
of massive leptons of a given third component of isospin are obtained.
To sum up, in the language of the Clifford algebra of nonrelativistic phase space
the lepton mass term corresponds to element $G_0$ (or $G^{\dagger}_0$). 

\subsection{Quark mass term}
We now want to transform the lepton mass element into the quark mass element.
In \cite{Zen2008,APPB2007b} it was shown that the transformation
from lepton to quark \# 2 and vice versa is obtained by choosing 
$\phi =+\pi/2$
in the genuine phase-space rotation operator
\begin{eqnarray}
{\cal{R}}_{02,\pm}(\phi)&=& e^{+i\phi F^{}_{\pm 2}}.
\end{eqnarray}
 Since this lepton-to-quark transformation constitutes an important part of the
present paper, we repeat this calculation for the relevant
projection operators.\\

In order to study the action of  
${\cal{R}}_{02,\pm} \equiv {\cal{R}}_{02,\pm}(+\pi/2)$, we first note that
$ (F_{\pm n})^3=F_{\pm n}$. Therefore
\begin{eqnarray}
{\cal{R}}_{02,\pm}&=&1+iF_{\pm 2}-(F_{\pm 2})^2.
\end{eqnarray}
Then, we calculate 
\begin{eqnarray}
F_{\pm 2}{\cal{Y}}_{k}&=&
\mp i(1-\delta_{2k})F_{\mp 2}B-\delta_{2k}F_{\pm 2},\nonumber\\
(F_{\pm 2})^2{\cal{Y}}_{k}&=&
\frac{1}{2}(1-\delta_{2k})({\cal{Y}}-{\cal{Y}}_2)
-\frac{1}{2}\delta_{2k}(1-{\cal{Y}}_2),
\end{eqnarray}
and find that:
\begin{eqnarray}
{\cal{R}}_{02,\pm}{\cal{Y}}_1{\cal{R}}_{02,\pm}^{-1}&=&-{\cal{Y}}_3,\nonumber\\
{\cal{R}}_{02,\pm}{\cal{Y}}_2{\cal{R}}_{02,\pm}^{-1}&=&+{\cal{Y}}_2,\nonumber\\
{\cal{R}}_{02,\pm}{\cal{Y}}_3{\cal{R}}_{02,\pm}^{-1}&=&-{\cal{Y}}_1.
\end{eqnarray}
Thus, the projection operators transform as
\begin{eqnarray}
{\cal{R}}_{02,\pm}Y_{+\frac{1}{3},1}{\cal{R}}_{02,\pm}^{-1}&=&Y_{+\frac{1}{3},1},\nonumber\\
{\cal{R}}_{02,\pm}Y_{+\frac{1}{3},2}{\cal{R}}_{02,\pm}^{-1}&=&Y_{-1},\nonumber\\
{\cal{R}}_{02,\pm}Y_{+\frac{1}{3},3}{\cal{R}}_{02,\pm}^{-1}&=&Y_{+\frac{1}{3},3},\nonumber\\
\label{projtransf}
{\cal{R}}_{02,\pm}Y_{-1}{\cal{R}}_{02,\pm}^{-1}&=&Y_{+\frac{1}{3},2},
\end{eqnarray}
i.e. quark \#2 and lepton are interchanged.\\

In order to know how lepton mass element $G_0$ transforms, we need to know
the individual actions of $F_{\pm 2}$ and $(F_{\pm 2})^2$
 on various odd elements of Clifford algebra. The relevant
formulas are gathered in Appendix B. Using Eqs (\ref{F+-2onodd}) we get:
\begin{eqnarray}
{\cal{R}}_{02,\pm}G_0{\cal{R}}_{02,\pm}^{-1}&=&\mp G_2,\nonumber\\
{\cal{R}}_{02,\pm}G_1{\cal{R}}_{02,\pm}^{-1}&=&+G_1,\nonumber\\
{\cal{R}}_{02,\pm}G_2{\cal{R}}_{02,\pm}^{-1}&=&\mp G_0,\nonumber\\
\label{R02+-Gk}
{\cal{R}}_{02,\pm}G_3{\cal{R}}_{02,\pm}^{-1}&=&+G_3.
\end{eqnarray}
Thus, the lepton mass element $G_0$ is transformed by the
${\cal{R}}_{02,\mp}$- induced transformations
into $\pm G_{2}$ (and vice versa). 
Consequently, element $G_2$, a member of $SU(3)$ sextet, should correspond to the mass
term of quark \# 2.
Just like in the case of the commutation relations for a quark of a given colour (see \cite{Zen2008,APPB2007b}), this mass term is not
rotationally invariant.
We shall comment on this lack of rotational invariance somewhat
later.

The difference between ${\cal{R}}_{02,-}$- and 
${\cal{R}}_{02,+}$-induced transformations is a $U(1)\otimes SU(3)$ 
phase factor ${\cal{F}}_{\{13\}}$:
\begin{equation}
\label{gauge}
{\cal{F}}_{\{13\}}{{\cal{R}}}_{02,-}\equiv \exp\left({i\frac{\pi}{2}F_{\{13\}}}\right){{\cal{R}}}_{02,-}
=\left(1+iF_{\{13\}}-F^2_{\{13\}}\right){{\cal{R}}}_{02,-}
={{\cal{R}}}_{02,+},
\end{equation}
where
\begin{equation}
F_{\{13\}}\equiv \frac{1}{2}F^{}_3-\frac{1}{2\sqrt{3}}F^{}_8+
\frac{1}{3}R^{} 
=H_{11}+H_{33}.
\end{equation}
This $U(1)\otimes SU(3)$ factor keeps commutation
relations invariant. 
Under ${\cal{F}}_{\{13\}}$-induced transformations, elements $G_0$ and $G_2$ change signs,
while $G_1$ and $G_3$ stay invariant. 
The total reflection, i.e. in particular $G_0,G_k \to -G_0,-G_k$
($k=1,2,3$),
is obtained through the consecutive action of 
${\cal{F}}_{\{13\}}$-, ${\cal{F}}_{\{23\}}$-, and ${\cal{F}}_{\{12\}}$- induced
transformations.

\subsection{SO(3) scalars}
Individual elements corresponding to masses of coloured quarks, i.e.
\begin{equation}
G_k=G_{\{\underline{kk}\}}
\end{equation}
belong to the $SU(3)$ sextet.
When summed over quark colours, we obtain 
the trace of the symmetric matrix $G_{\{kl\}}$, i.e.:
\begin{eqnarray}
G \equiv G_{\{kk\}}&=&\pm {\cal{R}}_{0{k},\mp}G_0{\cal{R}}_{0{k},\mp}^{-1},
\end{eqnarray}
which is an SO(3) scalar. 
Thus, when added, the three rotationally-noninvariant individual quark mass elements
give an
SO(3)-inva\-riant overall mass element. This provides an example of quark
conspiration, a conjecture originally put forward in \cite{APPB2007}-\cite{Zen2006}.
Interestingly (especially for a Clifford algebra approach), the quark mass element
 appears to be a trace of a rank 2 {\it symmetric} tensor
in our 3D world.

The products of 
the quark mass elements $G$ and $G^{\dagger}$ are (using Eqs (\ref{GG}, \ref{GkGl})): 
\begin{eqnarray}
G^2=(G^{\dagger})^2&=&0,\nonumber\\
GG^{\dagger}=
G_nG^{\dagger}_n&=&
2Y^+_{+\frac{1}{3}},\nonumber\\
G^{\dagger}G=
G^{\dagger}_nG_n&=&
2Y^-_{+\frac{1}{3}},
\end{eqnarray}
mirroring the behaviour of the products of $G_0$ and $G_0^{\dagger}$ 
(Eq. (\ref{E0E0dagger})), but in the
$Y=+\frac{1}{3}$ subspace.
In particular, the latter two expressions, 
just like $G_0G^{\dagger}_0$ and $G^{\dagger}_0G_0$, 
are invariant under both rotations and reflections.

The odd element $I_+$, with which we started in Eq. (\ref{odd}),
is a linear superposition of $SU(3)$ singlet ($G_0$)
and sextet ($G_{\{\underline{kk}\}}$) terms:
\begin{equation}
I_+=G_0+G,
\end{equation}
with $G_0$ and $G$ proportional to weak isospin raising (lowering) operators in
lepton and quark subspaces respectively. The explicit form of $G$ is given in
Appendix A.

\subsection{Relation to phase space}
By direct calculation using Eqs (\ref{F+n},\ref{F-n}),  
or with the help of Appendix B and Eq. (\ref{W+V+U}), 
we find:
\begin{eqnarray}
\label{RnminusC}
\tilde{C}_k={\cal{R}}_{0\underline{n},-}C_k{\cal{R}}_{0\underline{n},-}^{-1}&=&
\delta_{k\underline{n}}C_{\underline{n}}+
\epsilon_{k\underline{n}m}C^{\dagger}_m,\\
\label{RnplusC}
\tilde{C}'_k={\cal{R}}_{0\underline{n},+}C_k{\cal{R}}_{0\underline{n},+}^{-1}&=&
\delta_{k\underline{n}}C_{\underline{n}}
+i\epsilon_{k\underline{n}m}C^{\dagger}_m.
\end{eqnarray}
Obviously, when acted upon by the appropriate $U(1)\otimes
SU(3) $
transformation (as in Eq.(\ref{gauge})), $\tilde{C}_k$
goes over into $\tilde{C}'_k$.\\

\subsubsection{${\cal{R}}_{0\underline{2},-}$-induced transformations}
For the ${\cal{R}}_{0\underline{2},-}$ transformations one gets
\begin{eqnarray}
\tilde{A}_k&=&
=A_2\delta_{2k}+\epsilon_{2kn}A_n,\nonumber\\
\tilde{B}_k&=&
=B_2\delta_{2k}-\epsilon_{2kn}B_n,
\end{eqnarray}
and similarly for the related transformations of momenta and positions:
\begin{eqnarray}
\tilde{p}_k&=&p_2\delta_{2k}+\epsilon_{2kn}p_n,\nonumber\\
\tilde{x}_k&=&x_2\delta_{2k}-\epsilon_{2kn}x_n.
\end{eqnarray}
Then:
\begin{eqnarray}
{\bf p}^2&\to&\tilde{{\bf p}}^2={\bf p}^2,\nonumber\\
{\bf x}^2&\to&\tilde{{\bf x}}^2={\bf x}^2.
\end{eqnarray}
This transformation is somewhat similar to ordinary rotations in that 
it does not ``mix" the physical momentum and
physical position spaces. 
The difference is that rotation in each subspace proceeds here in the sense 
opposite to that in the
other subspace. 

\subsubsection{${\cal{R}}_{0\underline{2},+}$-induced transformations}
For the ${\cal{R}}_{0\underline{2},+}$ transformations one has
\begin{eqnarray}
\label{Atransformed}
\tilde{A}'_k
&=&\delta_{2k}A_2-\epsilon_{2kn}B_n,\nonumber\\
\label{Btransformed}
\tilde{B}'_k
&=&\delta_{2k}B_2-\epsilon_{2kn}A_n.
\end{eqnarray}
and similarly for the related transformations of momenta and positions:
\begin{eqnarray}
\tilde{p}'_k
&=&\delta_{2k}p_2-\epsilon_{2kn}x_n,\nonumber\\
\label{xpR02+}
\tilde{x}'_k
&=&\delta_{2k}x_2-\epsilon_{2kn}p_n.
\end{eqnarray}
Thus:
\begin{eqnarray}
{\bf p}^2 & \to & ({\bf \tilde{p}}')^2=x^2_1+p^2_2+x^2_3,\nonumber\\
{\bf x}^2 & \to & ({\bf \tilde{x}}')^2=p^2_1+x^2_2+p^2_3.
\end{eqnarray}
The above behaviour is different from 
the ${\cal{R}}_{02,-}$ case.
In the ${\cal{R}}_{02,+}$ case it is only its double application that leads
to the preservation of ${\bf p}^2$ and ${\bf x}^2$.

\subsubsection{Permutations of phase-space variables}
Connection between the above two genuine $SU(4)$ transformations
 is given by the $U(1)\otimes SU(3)$ phase factor of Eq. (\ref{gauge}).
This factor
 changes some position coordinates into momenta and vice versa
while keeping commutation relations invariant.
Thus, it violates the association between e.g. lepton mass and momentum.
Conseqently, it seems natural to suppose that, when talking about the concept of mass,
 the $U(1) \otimes
SU(3)$ freedom provided by ${\cal{F}}_{\{13\}}$  must be restricted
(as argued earlier), and the choice
between ${\cal{R}}_{02,\pm}$ limited to one possibility.
I think that the appropriate choice is provided by
 ${\cal{R}}_{02,+}$, which does not
keep the distinction between physical momenta and positions and may be considered
 more
``primitive" than the ${\cal{R}}_{02,-}$. As discussed in \cite{APPB2007b},
this choice corresponds to the 
third alternative (obtained from Eq.(\ref{xpR02+}) via ordinary 3D rotation by $\pi/2$ 
around the second axis)
of the following four choices
 for the ``generalized momenta" $P_k$ and
``generalized positions" $X_k$:
\begin{eqnarray}
\left[\begin{array}{c}
P_1,P_2,P_3\\
X_1,X_2,X_3
\end{array}
\right]&=& 
\left[\begin{array}{c}p_1,p_2,p_3\\ x_1,x_2,x_3\end{array}\right],
\left[\begin{array}{c}p_1,x_2,x_3\\ x_1,p_2,p_3\end{array}\right],
\left[\begin{array}{c}x_1,p_2,x_3\\ p_1,x_2,p_3\end{array}\right],
\left[\begin{array}{c}x_1,x_2,p_3\\ p_1,p_2,x_3\end{array}\right],
\label{genPgenX}
\end{eqnarray}
which are possible when pairs of permutations 
$(x_k,x_l) \leftrightarrow (p_k,p_l)$ are admitted. 
Such pairs of permutations 
keep an odd number of $p_k$'s in
${\bf P}$ and $x_k$'s in ${\bf X}$, and 
never permit a complete interchange ${\bf P} \leftrightarrow {\bf X}$.
 Thus, some distinction
between physical momentum and physical position is still preserved.
When expressed in terms of
$P_k$ and $X_l$, the 
commutation relations for each choice are then identical:
\begin{eqnarray}
[P_k,P_l]&=&0,\nonumber\\
{}[X_k,X_l]&=&0,\nonumber\\
\label{PX}
{}[X_k,P_l]&=&i\delta_{kl}.
\end{eqnarray}
Alternatively, one may replace $i$ with $-i$, which is equivalent to (\ref{PX}) after
genuine reflections in phase space are performed (e.g. ${\bf P} \to {\bf P}$, 
${\bf X} \to -{\bf X}$). 
The above commutation formulas are invariant under ``generalized rotations", 
appropriately and independently defined
for each of the four alternatives of Eq. (\ref{genPgenX}).

In my opinion, the emergence of a mass element which 
- for a quark of given colour - is not
rotationally invariant (and is associated with rotationally noninvariant
generalized momenta which  involve components of position) consitutes an asset of the approach,
and should be welcome in view of the unobservability 
of free quarks.
An object which does not satisfy the standard relation between
energy, mass, and momentum clearly cannot be seen as a "free particle".
This string-like idea on the origin of quark unobservability may coexist
with the standard description of strong interactions 
in terms of nonabelian $SU(3)$ gauge theory, which
is viewed here as an emergent theory,
 tested at small distances only (i.e. at large momenta when standard quark masses
should be neglected).

In fact, the application to quarks of the standard concept of mass leads to
conceptual difficulties, discussed at some length in \cite{Zen2006}.
These difficulties are related to the use of {\em free} solutions of the
Dirac equation (i.e. by setting $\displaystyle{\not}p\psi=m\psi$) 
for {\em confined} quarks, a standard procedure used e.g. when quark ``masses" are
``extracted" from the hadron-level 
data\footnote{It is conceptually consistent to consider quark mass 
as a parameter introduced in a standard (lepton-like) way into the underlying SM
lagrangian, and to determine this parameter
 from the observed hadron masses after the latter are obtained from
 a {\em full nonperturbative QCD calculation}. 
The problem is that this is {\em not} the way in which quark masses 
(listed e.g. by Particle Data Group) are extracted from the data.}.
In fact, precisely such a 
use of {free} quark solutions 
for {confined} quarks
has led to predictions for weak
radiative hyperon decays \cite{WRHD} which disagreed with experiment in a dramatic way.
The problem with the standard approach to quark mass is therefore not only conceptual.
So far, the case of weak radiative hyperon decays has been solved only by 
a departure from the use of the Dirac equation for quarks \cite{ZenWRHDsol},
and a treatment of quarks at the level of current algebra 
symmetries \cite{GM,CA}, with
the concept of quark mass simply avoided.

\section{Conclusions and Outlook}
In this paper the Clifford algebra of nonrelativistic phase space 
is discussed in some detail.
We have identified the element which may be associated with lepton mass and
transformed it to the quark sector.
The resulting individual quark mass element appears to be rotationally noninvariant.
The total quark mass term was then obtained as a sum of three individual quark mass
elements.
In this term
the individual rotationally-noninvariant contributions 
from three coloured quarks 
are combined in a rotationally invariant way,
in line with the expectations 
that quarks should conspire to yield rotationally covariant structures.

We have discussed one-particle system only. 
A further development of our approach should be presumably concerned 
with the description of nonrelativistic composite systems built of quarks.  
Hopefully, the idea of quark conspiracy could then be shown to work for such systems as well.
The question of a connection between total spin and mass should also be addressed.

The issues related to the emergence of a continuum (space, time, etc.), its
symmetries, and in particular an extension of the whole scheme to link it to
special relativity and accommodate gauge interactions, are clearly very
important.
In my opinion, however, they should be addressed only if a successful nonrelativistic 
treatment
of composite systems and higher $SO(3)$ representations is
developed.\\

\newpage

\renewcommand{\theequation}{A-\arabic{equation}}
\setcounter{equation}{0}
\section*{APPENDIX A - Explicit expressions}
\subsection*{Even elements}
When summed over $I_3=\pm \frac{1}{2}$, the even elements have the following explicit forms
\begin{eqnarray}
H^{}_{nk}&=&
\frac{1}{4}(\sigma_n\otimes\sigma_k+\sigma_k\otimes\sigma_n)\otimes \sigma_3
-\frac{i}{4}\epsilon_{nkm}(\sigma_m\otimes\sigma_0+\sigma_0\otimes\sigma_m)
\otimes\sigma_0,\nonumber\\
H^{}_{n0}&=&-\frac{i}{4}\left[
(\sigma_0\otimes\sigma_n-\sigma_n\otimes\sigma_0)\otimes\sigma_0
+i\,\epsilon_{nkl}\,\sigma_k\otimes\sigma_l\otimes\sigma_3
\right],\nonumber\\
H^{}_{0n}&=&+\frac{i}{4}\left[
(\sigma_0\otimes\sigma_n-\sigma_n\otimes\sigma_0)\otimes\sigma_0
-i\,\epsilon_{nkl}\,\sigma_k\otimes\sigma_l\otimes\sigma_3
\right],\nonumber\\
1&=&\sigma_0\otimes\sigma_0\otimes\sigma_0.
\end{eqnarray}
The projection operators correspond to the specific combinations:
\begin{eqnarray}
Y^{\pm}_{-1}&=&\frac{1}{4}(1-\sigma_m \otimes \sigma_m)\otimes 
\frac{\sigma_0\pm\sigma_3}{2},\nonumber\\
Y^{\pm}_{+\frac{1}{3}}&=&\frac{1}{4}(3+\sigma_m \otimes\sigma_m)\otimes 
\frac{\sigma_0\pm\sigma_3}{2}.
\end{eqnarray}

\subsection*{Odd elements}
The 16 elements proportional to $\sigma_1+i\sigma_2$ are
(in the order: SU(3) singlet, SU(3) antitriplet, two SU(3) triplets, 
SU(3) sextet):
\begin{eqnarray}
G_0&=&\frac{1-y}{4} \otimes 
\frac{\sigma_1 + i \sigma_2}{\sqrt{2}},
\nonumber\\
U^{\dagger}_k&=&-\frac{1}{{2}}(\sigma_0 \otimes \sigma_k+\sigma_k \otimes \sigma_0)
\otimes \frac{\sigma_1+i\sigma_2}{\sqrt{2}},\nonumber\\
V_k&=&\frac{1}{{4}}[\,(\sigma_0 \otimes \sigma_k-\sigma_k \otimes
\sigma_0)-i\epsilon_{kmn}\sigma_m\otimes\sigma_n]
\otimes \frac{\sigma_1+i\sigma_2}{\sqrt{2}}, \nonumber\\
W_k&=&\frac{1}{{4}}[\,(\sigma_0 \otimes \sigma_k-\sigma_k \otimes
\sigma_0)+i\epsilon_{kmn}\sigma_m\otimes\sigma_n]
\otimes \frac{\sigma_1+i\sigma_2}{\sqrt{2}},\nonumber\\
G_{\{kl\}}&=&\frac{1}{4}\left[\,
\delta_{kl}(\sigma_0\otimes\sigma_0+\sigma_m\otimes\sigma_m)-
(\sigma_k\otimes\sigma_l+\sigma_l\otimes\sigma_k)\right]
\otimes\frac{\sigma_1+i\sigma_2}{\sqrt{2}}.
\end{eqnarray}
The diagonal elements $G_{\{\underline{kk}\}}$ sum up to:
\begin{eqnarray}
G&=&G_{\{kk\}}=\sum_kG_k=
\frac{3+y}{4}\otimes \frac{\sigma_1+i\sigma_2}{\sqrt{2}},
\end{eqnarray}
with
\begin{eqnarray}
G_k&=&
=\frac{1}{4}(1+y-2y_k)\otimes 
\frac{\sigma_1+i\sigma_2}{\sqrt{2}}.
\label{AppEn}
\end{eqnarray}
Analogous expressions hold for 16 hermitian-conjugated
elements proportional to $\sigma_1-i\sigma_2$.

\renewcommand{\theequation}{B-\arabic{equation}}
\setcounter{equation}{0}
\section*{APPENDIX B - 
Products of elements}
\subsection*{Odd-odd}
\subsubsection*{{Triplet-triplet}}
One finds
\begin{eqnarray}
W_kW_l=V_kV_l=U_kU_l=V_kW_l=W_kV_l=V_kU_l=U_kW_l&=&0,
\end{eqnarray} 
and
\begin{eqnarray}
W_kU_l&=&
2\epsilon_{mkl}Y^+_{-1}H^{}_{m0}Y^+_{+\frac{1}{3}}=2\epsilon_{mkl}H^+_{m0},
\nonumber\\
U_kV_l&=&
2\epsilon_{mkl}Y^-_{+\frac{1}{3}}H^{}_{m0}Y^-_{-1}=2\epsilon_{mkl}H^-_{m0},
\end{eqnarray}
with analogous formulae for the h.c. expressions.
Furthermore
\begin{eqnarray}
W_kV_l^{\dagger}=W^{\dagger}_kV_l=V_kU^{\dagger}_l=
V^{\dagger}_kU_l=U_kW_l^{\dagger}=U^{\dagger}_kW_l&=&0,
\end{eqnarray}
together with h.c. relations. Using Eqs (\ref{WVUprojections})
we have
\begin{eqnarray}
W_kW_l^{\dagger}&=&
Y^+_{-1}\{C_k,C_l^{\dagger}\}Y^+_{-1}
=2\delta_{kl}Y^+_{-1},\nonumber\\
\label{WWdagVdagV1}
V^{\dagger}_kV_l&=&
Y^-_{-1}\{C_k^{\dagger},C_l\}Y^-_{-1}
=2\delta_{kl}Y^-_{-1},\nonumber\\
V_kV^{\dagger}_l+U^{\dagger}_lU_k&=&
Y^+_{+\frac{1}{3}}\{C_k,C^{\dagger}_l\}
Y^+_{+\frac{1}{3}}
=2\delta_{kl}Y^+_{+\frac{1}{3}},\nonumber\\
\label{WWdagVdagV2}
W^{\dagger}_kW_l+U_lU^{\dagger}_k&=&
Y^-_{+\frac{1}{3}}\{
C_k^{\dagger},C_l\}Y^-_{+\frac{1}{3}}
=2\delta_{kl}Y^-_{+\frac{1}{3}},\nonumber\\
V_kV_l^{\dagger}-U^{\dagger}_lU_k&=&
Y^+_{+\frac{1}{3}}[C_k,C^{\dagger}_l]
Y^+_{+\frac{1}{3}}=-4H^+_{kl}-2\delta_{kl}Y^+_{-1},\nonumber\\
U_kU^{\dagger}_l-W^{\dagger}_lW_k&=&
Y^-_{+\frac{1}{3}}[
C_k,C_l^{\dagger}]Y^-_{+\frac{1}{3}}=-4H^-_{kl}+2\delta_{kl}Y^-_{-1}.
\end{eqnarray}

\subsubsection*{Triplet-singlet}
One finds 
\begin{eqnarray}
W_kG_0=V_kG_0=U_kG_0=C_kG_0&=0=&G_0C_k=G_0W_k=G_0V_k=G_0U_k,\nonumber\\
V^{\dagger}_kG_0=U^{\dagger}_kG_0&=0=&G_0W^{\dagger}_k=G_0U^{\dagger}_k,
\end{eqnarray}
and
\begin{eqnarray}
W^{\dagger}_kG_0=-iC^{\dagger}_kG_0&=&2iY^-_{+\frac{1}{3}}H_{k0}=
2iH_{k0}Y^-_{-1}=2iH^-_{k0},
\nonumber\\
G_0V^{\dagger}_k=-iG_0C_k^{\dagger}&=&2iY^+_{-1}H_{k0}=
2iH_{k0}Y^+_{+\frac{1}{3}}=2iH^+_{k0},
\end{eqnarray}
and similarly for the h.c. expressions.

\subsubsection*{Singlet-singlet}
\begin{eqnarray}
(G_{0})^2=(G^{\dagger}_{0})^2&=&0,\nonumber\\
G_{0}G^{\dagger}_{0}&=&2Y^+_{-1},\nonumber\\
\label{G0G0+}
{}G^{\dagger}_{0}G_{0}&=&2Y^-_{-1}.
\end{eqnarray}

\subsubsection*{Sextet-sextet}
From the isospin structure of sextet elements
one has (for any $k$, $l$, $m$, $n$):
\begin{equation}
\label{GG}
G_{\{kl\}}G_{\{mn\}}=0=G^{\dagger}_{\{kl\}}G^{\dagger}_{\{mn\}}.
\end{equation}
The products $G_{\{kl\}}G_{\{mn\}}^{\dagger}$
and $G^{\dagger}_{\{kl\}}G_{\{mn\}}$ belong to two
separate (isospin) subspaces,
with the general formulas (valid for any $k$, $l$, $m$, $n$) being:
\begin{eqnarray}
G_{\{kl\}}G^{\dagger}_{\{mn\}}&=&\frac{1}{2}\left\{
\left(\frac{1}{2}+H^{}_{jj}\right)
(\delta_{km}\delta_{ln}+\delta_{kn}\delta_{lm})\right.\nonumber\\
&&-\left.\left(\delta_{km}H^{}_{ln}+\delta_{kn}H^{}_{lm}
+\delta_{lm}H^{}_{kn}+\delta_{ln}H^{}_{km}
\right)\right\}I_{+\frac{1}{2}},
\end{eqnarray}
and
\begin{eqnarray}
G^{\dagger}_{\{kl\}}G_{\{mn\}}&=&\frac{1}{2}\left\{
\left(\frac{1}{2}-H^{}_{jj}\right)
(\delta_{mk}\delta_{nl}+\delta_{nk}\delta_{ml})\right.\nonumber\\
&&+\left.\left(\delta_{mk}H^{}_{nl}+\delta_{nk}H^{}_{ml}
+\delta_{ml}H^{}_{nk}+\delta_{nl}H^{}_{mk}
\right)\right\}I_{-\frac{1}{2}}.
\end{eqnarray}
Partial cases of the above formulas are (for $k \ne l$ !):
\begin{eqnarray}
G_{\{\underline{kl}\}}G^{\dagger}_{\{\underline{kl}\}}=
G_{\{\underline{kl}\}}G^{\dagger}_{\{\underline{lk}\}}
&=&\frac{1}{2}(Y^+_{+\frac{1}{3},k}+Y^+_{+\frac{1}{3},l}),\nonumber\\
G^{\dagger}_{\{\underline{kl}\}}G_{\{\underline{kl}\}}=
G^{\dagger}_{\{\underline{kl}\}}G_{\{\underline{lk}\}}
&=&\frac{1}{2}(Y^-_{+\frac{1}{3},k}+Y^-_{+\frac{1}{3},l}),
\end{eqnarray}
and for any $k,m$:
\begin{eqnarray}
G_kG^{\dagger}_m=G_{\{\underline{kk}\}}G^{\dagger}_{\{\underline{mm}\}}&=&
2\delta_{k\underline{m}}Y^+_{+\frac{1}{3},\underline{m}},\nonumber\\
\label{GkGl}
\label{GkkGmm}
G^{\dagger}_kG_m=G^{\dagger}_{\{\underline{kk}\}}G_{\{\underline{mm}\}}&=&
2\delta_{k\underline{m}}Y^-_{+\frac{1}{3},\underline{m}}.
\end{eqnarray}
Furthermore, two more typical cases are (for $l\ne k,n$ and $n \ne k$)
\begin{eqnarray}
G_{\{k\underline{l}\}}G^{\dagger}_{\{\underline{l}n\}}&=&
-\frac{1}{2}H^{+}_{kn},\nonumber\\
G^{\dagger}_{\{k\underline{l}\}}G_{\{\underline{l}n\}}&=&
+\frac{1}{2}H^{-}_{nk},
\end{eqnarray}
and (for $k \ne n$)
\begin{eqnarray}
G_{\{\underline{kk}\}}G^{\dagger}_{\{\underline{k}n\}}
=G_{\{\underline{n}k\}}G^{\dagger}_{\{\underline{nn}\}}&=&
-H^{+}_{kn},\nonumber\\
G^{\dagger}_{\{\underline{kk}\}}G_{\{\underline{k}n\}}
=G^{\dagger}_{\{\underline{n}k\}}G_{\{\underline{nn}\}}&=&
+H^{-}_{nk}.
\end{eqnarray}
All other types of products yield zero.
To summarize, all expressions on the r.h.s of the above equations contain only
the unit operator and
 the generators of $U(1)\otimes SU(3)$ projected upon subspaces with
 $I_{3}=\pm\frac{1}{2}$.

\subsubsection*{Sextet-triplet}
Of all 24 products of $G_{\{kl\}}$ and $G_{\{kl\}}^{\dagger}$ with
$U_k$, $V_k$, $W_k$, $U^{\dagger}_k$, $V^{\dagger}_k$, and $W^{\dagger}_k$, 
only eight are nonzero, i.e.
\begin{eqnarray}
G_{\{kl\}}U_n&=&+i(\epsilon_{mln}H^+_{km}+\epsilon_{mkn}H^+_{lm}),\nonumber\\
G^{\dagger}_{\{kl\}}U^{\dagger}_n&=&
-i(\epsilon_{mln}H^-_{mk}+\epsilon_{mkn}H^-_{ml}),\nonumber\\
G_{\{kl\}}W^{\dagger}_n&=&-i(\delta_{kn}H^+_{0l}+\delta_{ln}H^+_{0k}),\nonumber\\
G^{\dagger}_{\{kl\}}V_n&=&+i(\delta_{kn}H^-_{l0}+\delta_{ln}H^-_{k0}),
\end{eqnarray}
and their h.c. versions.

\subsubsection*{Sextet-singlet}
Here all products are zero because sextet and singlet elements
correspond to diferrent values of hypercharge.\\

\subsection*{Even-odd}
\subsubsection*{Shift operators of $U(1) \otimes SU(3)$ and odd elements}
The nonzero products of the  $U(1)\otimes SU(3)$ shift operators with the odd elements are:
\begin{eqnarray}
H^+_{kl}U_n^{\dagger}&=&+\frac{1}{2}\delta_{kn}U^{\dagger}_l-
i\epsilon_{mnl}G_{\{mk\}},\nonumber\\
H^+_{kl}V_n&=&+\frac{1}{2}\delta_{kl}V_n-\delta_{nl}V_k,\nonumber\\
H^+_{kl}W_n&=&-\frac{1}{2}\delta_{kl}W_n^{},\nonumber\\
H^+_{kl}G_0&=&-\frac{1}{2}\delta_{kl}G_0,\nonumber\\
H^+_{kl}G_{\{mn\}}&=&-\frac{i}{4}(\delta_{ml}\epsilon_{nkr}
+\delta_{nl}\epsilon_{mkr})U^{\dagger}_r+\nonumber\\
&&-\frac{1}{2}(\delta_{ml}G_{\{nk\}}+\delta_{nl}G_{\{mk\}}-\delta_{kl}G_{\{mn\}}),
\end{eqnarray}
and
\begin{eqnarray}
H^-_{kl}U_n^{}&=&-\frac{1}{2}\delta_{ln}U^{}_k+
i\epsilon_{mnk}G_{\{ml\}}^{\dagger},\nonumber\\
H^-_{kl}V_n^{\dagger}&=&+\frac{1}{2}\delta_{kl}V_n^{\dagger},\nonumber\\
H^-_{kl}W_n^{\dagger}&=&-\frac{1}{2}\delta_{kl}W_n^{\dagger}+\delta_{kn}W^{\dagger}_l,\nonumber\\
H^-_{kl}G_0^{\dagger}&=&+\frac{1}{2}\delta_{kl}G_0^{\dagger},\nonumber\\
H^-_{kl}G_{\{mn\}}^{\dagger}&=&+\frac{i}{4}(\delta_{mk}\epsilon_{nlr}
+\delta_{nk}\epsilon_{mlr})U^{}_r+\nonumber\\
&&+\frac{1}{2}(\delta_{mk}G_{\{nl\}}^{\dagger}+\delta_{nk}G_{\{ml\}}^{\dagger}
-\delta_{kl}G_{\{mn\}}^{\dagger}).
\end{eqnarray}

\subsubsection*{Genuine generators of $SU(4)$ and odd elements}
The nonzero products of $H_{n0}$, $H_{0n}$ and $F^{}_{\pm n}$ 
with odd elements of Clifford algebra are:
\begin{eqnarray}
F^{}_{-n}V_k=+iF^{}_{+n}V_k=+iH_{n0}V_k&=&\delta_{nk}G_0,\nonumber\\
F^{}_{-n}V_k^{\dagger}=+iF^{}_{+n}V^{\dagger}_k=+iH_{n0}V^{\dagger}_k&=
&-\frac{i}{2}\epsilon_{nkl}U_l+G^{\dagger}_{\{nk\}},\nonumber\\
F^{}_{-n}W_k=-iF^{}_{+n}W_k=-iH_{0n}W_k&=
&-\frac{i}{2}\epsilon_{nkl}U^{\dagger}_l+G_{\{nk\}},\nonumber\\
F^{}_{-n}W^{\dagger}_k=-iF^{}_{+n}W^{\dagger}_k=-iH_{0n}W^{\dagger}_k&=
&\delta_{nk}G_0^{\dagger},\nonumber\\
F^{}_{-n}U_k=-iF^{}_{+n}U_k=-iH_{0n}U_k&=&-i\epsilon_{nkl}V^{\dagger}_l,\nonumber\\
F^{}_{-n}U^{\dagger}_k=+iF^{}_{+n}U^{\dagger}_k=+iH_{n0}U^{\dagger}_k&=
&-i\epsilon_{nkl}W_l,\nonumber\\
F^{}_{-n}G_0=-iF^{}_{+n}G_0=-iH_{0n}G_0&=&V_n,\nonumber\\
F^{}_{-n}G_0^{\dagger}=+iF^{}_{+n}G_0^{\dagger}=+iH_{n0}G^{\dagger}_0&=&W^{\dagger}_n,\nonumber\\
F^{}_{-n}G_{\{kl\}}=+iF^{}_{+n}G_{\{kl\}}=+iH_{n0}G_{\{kl\}}&=&
\frac{1}{2}\delta_{nk}W_l+\frac{1}{2}\delta_{nl}W_k,\nonumber\\
\label{F+-2onodd}
F^{}_{-n}G^{\dagger}_{\{kl\}}=-iF^{}_{+n}G^{\dagger}_{\{kl\}}=
-iH_{0n}G^{\dagger}_{\{kl\}}&=&
\frac{1}{2}\delta_{nk}V^{\dagger}_l+\frac{1}{2}\delta_{nl}V^{\dagger}_k.
\end{eqnarray}
For completeness, we also specify how $U_k$, $V_k$, and $W_k$ transform under
${\cal{R}}_{02,\pm}$-induced transformations:
\begin{eqnarray}
{\cal{R}}_{02,-}V_k{\cal{R}}_{02,-}^{-1}&=&\delta_{2k}W_2+
\frac{1}{2}\,\epsilon_{2kl}\,U^{\dagger}_l-{i}(1-\delta_{2k})G_{\{2k\}},
\nonumber\\
{\cal{R}}_{02,-}W_k{\cal{R}}_{02,-}^{-1}&=&\delta_{2k}V_2+
\frac{1}{2}\,\epsilon_{2kl}\,U^{\dagger}_l+{i}(1-\delta_{2k})G_{\{2k\}},
\nonumber\\
{\cal{R}}_{02,-}U_k{\cal{R}}_{02,-}^{-1}&=
&\delta_{2k}U_2+\epsilon_{2kl}\,(V^{\dagger}_l+W^{\dagger}_l),
\end{eqnarray}
and
\begin{eqnarray}
{\cal{R}}_{02,+}V_k{\cal{R}}_{02,+}^{-1}&=&
\delta_{k2}W_2+\frac{i}{2}\,\epsilon_{2kl}\,U^{\dagger}_l+(1-\delta_{k2})
G_{\{2k\}},
\nonumber\\
{\cal{R}}_{02,+}W_k{\cal{R}}_{02,+}^{-1}&=&
\delta_{k2}V_2+\frac{i}{2}\,\epsilon_{2kl}\,U^{\dagger}_l-(1-\delta_{k2})
G_{\{2k\}},\nonumber\\
{\cal{R}}_{02,+}U_k{\cal{R}}_{02,+}^{-1}&=&
\delta_{k2}U_2+
{i}\,\epsilon_{2kl}\,(V^{\dagger}_l+W^{\dagger}_l),
\end{eqnarray}
with analogous equations for the hermitean conjugates.\\

Under 
${\cal{R}}_{02,\pm}$-induced transformations, the off-diagonal ($k \ne l$)
elements $G_{\{kl\}}$  transform as:
\begin{eqnarray}
{\cal{R}}_{02,-}G_{\{12\}}{\cal{R}}_{02,-}^{-1}=
+i\,{\cal{R}}_{02,+}G_{\{12\}}{\cal{R}}_{02,+}^{-1}&=&+\frac{i}{2}(W_1-V_1),
\nonumber\\
{\cal{R}}_{02,-}G_{\{13\}}{\cal{R}}_{02,-}^{-1}=
\,{\cal{R}}_{02,+}G_{\{13\}}{\cal{R}}_{02,+}^{-1}&=&G_{\{13\}},
\nonumber\\
{\cal{R}}_{02,-}G_{\{23\}}{\cal{R}}_{02,-}^{-1}=
+i\,{\cal{R}}_{02,+}G_{\{23\}}{\cal{R}}_{02,+}^{-1}&=&+\frac{i}{2}(W_3-V_3).
\end{eqnarray}
The corresponding formulas for $G_{\{\underline{kk}\}}$ are given in Eq. (\ref{R02+-Gk}).

\vfill

\vfill

\end{document}